\documentclass[11pt]{article}

\usepackage{graphicx, amssymb, latexsym, amsfonts, amsmath, lscape, amscd,
amsthm, color, epsfig, mathrsfs, tikz, enumerate, multirow, mathrsfs}
\usepackage{tikz,pgf,tkz-graph}
\usetikzlibrary{shapes.geometric,calc,shapes,fit,intersections,graphs,graphs.standard,decorations.pathreplacing,decorations.markings,positioning,arrows,shadows,arrows.meta,fadings,decorations.text}
\usetikzlibrary{arrows,automata}
\usetikzlibrary{graphs}
\usetikzlibrary{graphs.standard}
\tikzset{every state/.style={minimum size=0pt}}
\usepackage{booktabs}
\usetikzlibrary{hobby}
\usepackage[]{mdframed}
\tikzset{XOR/.style={draw,circle,append after command={
 [shorten >=\pgflinewidth, shorten <=\pgflinewidth,]
 (\tikzlastnode.north) edge (\tikzlastnode.south)
 (\tikzlastnode.east) edge (\tikzlastnode.west)
 }
 }
}

\usepackage[textsize=footnotesize,color=green!40]{todonotes}
\newtheorem{theorem}{Theorem}[section]

\newtheorem{corollary}[theorem]{Corollary}

\newtheorem{lemma}[theorem]{Lemma}

\newtheorem{algo}[theorem]{Algorithm}
\theoremstyle{definition}
\newtheorem{definition}[theorem]{Definition}
\newtheorem{observation}[theorem]{Observation}

\theoremstyle{definition}

\newcommand\DELETE[1]{}

\DeclareMathOperator{\MEG}{meg}

\usetikzlibrary{decorations.pathmorphing}

\usepackage{todonotes}

\begin{document}


 \title{\bf Algorithms and complexity for monitoring edge-geodetic sets in graphs
 }
\author{{\sc Florent Foucaud}$\,^{a}$, {\sc Clara Marcille}$\,^{b}$, 
{\sc 
R. B. Sandeep}$\,^{c}$, \\{\sc 
Sagnik Sen}$\,^{c}$, {\sc 
S Taruni}$\,^{d}$\\
\mbox{}\\
{\small $(a)$ Université Clermont Auvergne, CNRS, Clermont Auvergne INP,} \\ {\small Mines Saint-\'Etienne LIMOS, 63000 Clermont-Ferrand, France.}\\
{\small $(b)$ Univ. Bordeaux, CNRS, Bordeaux INP, LaBRI, UMR 5800,}\\ {\small F-33400, Talence, France.}\\
{\small $(c)$ Indian Institute of Technology Dharwad, India.}\\
{\small $(d)$ Centro de Modelamiento Matemático (CNRS IRL2807),}\\
{\small Universidad de Chile, Santiago, Chile.}\\
}

\date{\today}

\maketitle

\begin{abstract}
A monitoring edge-geodetic set of a graph is a subset $M$ of its vertices such that for every edge $e$ in the graph, deleting $e$ increases the distance between at least one pair of vertices in $M$. 
We study the following computational problem \textsc{MEG-set}: given a graph $G$ and an integer $k$, decide whether $G$ has a monitoring edge geodetic set of size at most $k$. We prove that the problem is NP-hard even for 2-apex 3-degenerate graphs improving a result by Haslegrave (Discrete Applied Mathematics 2023). Additionally, we prove that the problem cannot be solved in subexponential-time, assuming the Exponential-Time Hypothesis, even for 3-degenerate graphs. Further, we prove that the optimization version of the problem is APX-hard even for 4-degenerate graphs. Complementing these hardness results, we prove that the problem admits a polynomial-time algorithm for interval graphs, a fixed-parameter tractable algorithm for general graphs with clique-width plus diameter as the parameter, and a fixed-parameter tractable algorithm for chordal graphs with treewidth as the parameter. We also provide an approximation algorithm with factor $\ln m\cdot OPT$ and $\sqrt{n\ln m}$ for the optimization version of the problem, where $m$ is the number of edges, $n$ the number of vertices, and $OPT$ is the size of a minimum monitoring edge-geodetic set of the input graph.
\end{abstract}


\section{Introduction}

Given a graph $G$, a \emph{monitoring edge-geodetic set} of $G$, or simply, an \emph{MEG-set} of a graph $G$, is a vertex subset $M \subseteq V(G)$ that satisfies the following: given any edge $e$ in $G$, there exists $u, v \in M$ such that $e$ lies on all shortest paths between $u$ and $v$. When this is the case, we say that the vertices $u,v$ \emph{monitor} the edge $e$. The \emph{monitoring edge-geodetic number}, denoted by $\MEG(G)$, is the smallest size of an MEG-set of $G$. This concept was introduced in \cite{foucaud2023monitoring}. 

MEG-sets are motivated by applications in network monitoring. The vertices of the MEG-set represent distance probes in a network, that can measure the distance between each other. 
If a connection (edge) of the network fails, the distance between two probes increases; they able to detect the change in distance, thereby the fault in the network. 

MEG-sets are related to several similar parameters on network monitoring like geodetic number, edge-geodetic number, strong-edge geodetic number, distance-edge monitoring number to name a few~\cite{atici2003edge, chartrand2002geodetic, foucaud2022monitoring, harary1993geodetic,irvsivc2018strong,santhakumaran2007edge,yannakakis1981edge}. 

\subsection*{Overview of earlier works}

In the introductory paper on MEG-sets \cite{foucaud2023monitoring} (whose extended version is~\cite{foucaud2023monitoringfull}), the parameter $\MEG(G)$ was determined for various graph families such as trees, cycles, unicyclic graphs, complete graphs, complete multipartite graphs, hypercubes and grids. This parameter was further studied for Mycielskian graph classes~\cite{li2024monitoring}, line graphs \cite{bao2023monitoring}, Cartesian and strong products of graphs~\cite{haslegrave2023monitoring}, corona products~\cite{foucaud2023monitoringfull}, cluster, lexicographic product, direct product and the join operation~\cite{xu2024monitoring}. The effects of clique sum and subdivision on $\MEG(G)$ are also explored in \cite{foucaud2024bounds}. The relation between $\MEG(G)$ and other parameters of graphs has also been studied; in particular, an upper bound using the parameter feedback edge set number $f$ and the number $\ell$ of vertices of degree $1$ of the graph, was given in~\cite{foucaud2023monitoring,foucaud2023monitoringfull}. This was improved to $\MEG(G)\leq 3f+\ell+1$ (which is tight up to an additive factor of 1) in~\cite{CFH23}. The relation between $\MEG(G)$ and other various closely related geodetic-parameters is studied in~\cite{foucaud2024bounds}. 

Many interesting examples where $\MEG(G)$ is exactly equal to the order of graph $G$ are found in both~\cite{foucaud2023monitoring} and~\cite{haslegrave2023monitoring}. Subsequently, an important necessary and sufficient condition for when a vertex is part of all MEG-sets or no MEG-set of a graph was determined in~\cite{foucaud2024bounds}, leading to a characterization (and polynomial-time recognition algorithm) of graphs where every vertex is in every MEG-set. Using this condition, it was shown in~\cite{foucaud2024bounds} that the value of the parameter $\MEG(G)$ can be determined in polynomial time for cographs, well-partitioned chordal graphs (which includes block graphs and split graphs), and proper interval graphs, since in these graph classes, an optimal MEG-set is obtained by selecting all vertices of the graph except for its cut-vertices. 

Deciding whether $\MEG(G)\leq k$ for general graphs is proven to be NP-complete in~\cite{haslegrave2023monitoring}. This was further improved to hold even for the graphs of maximum degree at most~9 in~\cite{foucaud2023monitoringfull}. Recently, the inapproximability of finding minimum MEG-set within $0.5\ln{n}$ was shown in~\cite{bilo2024inapproximability}.

\subsection*{Our contributions}
This paper is based on some parts of a shorter version that was published in the proceedings of the CALDAM 2024 conference~\cite{FMMSST24}. The present paper contains the extended version of the computational complexity results of this conference version, with additional approximation results along with new exact algorithmic results. A separate paper~\cite{foucaud2024bounds} contains an extension of the structural results from~\cite{FMMSST24}, and the results from~\cite{foucaud2024bounds} are disjoint from those in the present paper.

The computational decision problem \textsc{MEG-set} is defined as follows.
\begin{mdframed}
\textsc{MEG-set} \\
\textbf{Instance:} A graph $G$, an integer $k$. \\
\textbf{Question:} Is it true that $\MEG(G) \leq k$? 
\end{mdframed}


\begin{itemize}
 \item[1.] In Section~\ref{sec:interval}, we present an exact polynomial-time algorithm for \textsc{MEG-set} for interval graphs, by providing a characterization of optimal MEG-sets of interval graphs using a result from~\cite{foucaud2024bounds}. This generalizes the result from~\cite{foucaud2024bounds} for proper interval graphs, for which the characterization is much simpler.

 \item[2.] In Section~\ref{sec FPT}, we show that \textsc{MEG-set} is fixed parameter tractable (FPT) for the parameter clique width + diameter by applying Courcelle's theorem. We also provide an explicit dynamic programming scheme on a tree decomposition of a chordal graph, showing that \textsc{MEG-set} is FPT for the parameter treewidth on chordal graphs. 

\item[3.] In Section~\ref{sec:approx}, we provide an approximation algorithm for the optimization version of \textsc{MEG-set}, that runs in polynomial time and has an approximation factor of both $\ln m\cdot OPT$ and $\sqrt{n\ln m}$, where $n$ is the number of vertices, $m$ is the number of edges, and $OPT$ is the size of a minimum MEG-set of the input graph. 

\item[4.] In Section~\ref{sec complexity}, We give an alternative proof for the NP-completeness of \textsc{MEG-set}. In fact, our proof is a strengthening of the previous proofs, as we show that \textsc{MEG-set} remains NP-complete even for the restricted class of $3$-degenerate, $2$-apex graphs. In addition to this, we show that the problem cannot be solved in subexponential-time, assuming the Exponential Time Hypothesis (ETH). We also show that there is no polynomial-time approximation scheme (assuming P $\neq$ NP) and the problem is APX-hard, even for 4-degenerate graphs.

\item[5.] Finally, we end the paper with concluding remarks in Section~\ref{sec conclusions}.
\end{itemize}

\section{Polynomial-time algorithm for interval graphs}\label{sec:interval}

It follows from results in~\cite{foucaud2024bounds} that \textsc{MEG-set} is solvable in polynomial time on proper interval graphs, split graphs, block graphs (more generally, on well-partitioned chordal graphs), and cographs. Indeed, for these graph classes, any optimal MEG-set consists of all vertices that are not cut-vertices. Although this property does not hold for interval graphs, we next use a characterization from~\cite{foucaud2024bounds} to show that \textsc{MEG-set} is solvable in polynomial time on interval graphs.

We begin with some preliminary notation. A graph is an \textit{interval graph} if we can find a set of intervals on the real line such that each vertex is assigned an interval and two vertices are adjacent if and only if their corresponding intervals intersect. Consider $G$ to be an interval graph. We define $I(a)$ to be the interval representation of a vertex $a$ in $G$. Let $l_a$ (resp., $r_a$) be the left (resp., right) end point of $I(a)$. Since $G$ is an interval graph, we can consider two total orderings on $G$, namely $<_l$ and $<_r$ as follows. 
We say $a <_{l} b$ (resp., $a <_r b$) if $l_a < l_b$ (resp., $r_a < r_b$) for any two vertices $a, b$ in $G$. Similarly, we denote $a=_l b$ (resp., $a=_r b$) when $l_a = l_b$ (resp., $r_a = r_b$).
Let $\ell^*$ be a vertex such that $I(\ell^*)$ is minimal with respect to $<_l$. Similarly, let $r^*$ be a vertex whose $I(r^*)$ is maximal with respect to $<_r$. Refer to~\cite{hsu1993simple} for an algorithm for finding an interval representation of an interval graph in polynomial time.

We recall an important necessary and sufficient condition for when a vertex is part of any MEG-set of a graph, proved in~\cite{foucaud2024bounds}. 

\begin{lemma}[\cite{foucaud2024bounds}]\label{lem property P}
 Let $G$ be a graph. A vertex $v\in V(G)$ is in every MEG-set of $G$ if and only if there exists $u \in N(v)$ such that any induced $2$-path $uvx$, where $x \in N(v)$, is part of a $4$-cycle.
\end{lemma}

If a vertex $v$ of a graph $G$ has the property mentioned in Lemma~\ref{lem property P}, then we say that $v$ is a \emph{mandatory vertex}, and call such a ``special'' vertex $u$ (that is, for any induced $2$-path $uvx$, where $x\in N(v)$, is part of a $4$-cycle) a \emph{support} of $v$. Observe that, if for no $x\in N(v)$, the path $uvx$ is an induced $2$-path, then $u$ is a support. This is, for instance, the case when $v$ is simplicial. For a vertex subset $S$ of $G$, we define the set, $$d_G(S) = \max \{ d_{G}(x,y) : x,y \in S \},$$
where $d_G(x,y)$ is the distance between two vertices $x$ and $y$ in the graph $G$. Using Lemma~\ref{lem property P}, we prove an interesting characterization in the form of the following corollary.

\begin{corollary}\label{cor property P}
 Let $v$ be a vertex in a graph $G$. If $v$ is a mandatory vertex in $G$, then $d_{G-v}(N(v)) \leq 4$. Moreover, if $G$ is an interval graph, then the converse is also true. 
\end{corollary}

\begin{proof}
Let $v$ be a mandatory vertex in $G$. Then by Lemma~\ref{lem property P} the set $N(v)$ contains a support $u$ of $v$. 
If $|N(v)|\leq 2$, then either $v$ is simplicial and we are done, or there exists $x_1\in N(v)$ distinct from $u$. By definition of a support, $d_{G-v}(x_1, u)\leq 2$ and we are done as well.
Let $x_1$ and $x_2$ be two vertices, distinct from $u$, of $N(v)$. Notice that, either $x_i$ is adjacent to $u$, or, by Lemma~\ref{lem property P}, has a $2$-path $uw_ix_i$, for $i \in \{1,2\}$ with $w_i\neq v$. In either case, 
we have $d_{G-v}(u, x_i) \leq 2$, and thus $d_{G-v}(x_i, x_j) \leq 4$.

Next let us assume that $G$ is an interval graph, and consider any interval representation of $G$. We want to show that the converse is also true in this case. 
Therefore, for some vertex $v$ in $G$, suppose that 
$d_{G-v}(N(v)) \leq 4$. We have to show that
$v$ is a mandatory vertex of $G$.

Notice that, if there exists another vertex $u$ satisfying $I(v) \subseteq I(u)$, 
then by Lemma~\ref{lem property P} $v$ is a mandatory vertex where $u$ is a support of $v$. Therefore, we can assume that there is no vertex $u$ in $G$ satisfying 
$I(v) \subseteq I(u)$. 
Let $a$ be minimal with respect to $<_r$ among all neighbors of $v$ and let $b$ be maximal with respect to $<_l$ among all neighbors of $v$. 
If $v_r \leq a_r$, then it implies that all intervals corresponding to each neighbor of $v$ intersect the point $a_r$, and hence $v$ is a mandatory vertex. Thus we may assume that 
$a_r < v_r$. Similarly, we may assume that $v_l < b_l$.

If $b_l \leq a_r$, then any neighbor $u$ of $v$ will contain the interval $[b_l,a_r]$ due to the definitions of $a$ and $b$. In particular, $a$ and $b$ will dominate 
$N(v)$, thus implying $v$ is a mandatory vertex where 
$a$ plays the role of its support. Thus we may assume that 
$a_r < b_l$. 

Therefore, for any $u \in N(v)$, $I(u) \cap [a_r, b_r] \neq \emptyset$. Let $P = au_1\cdots u_kb$ be a shortest path connecting $a$ and $b$ in $G-v$. Hence any $u \in N(v)$ must intersect an internal vertex of $P$. That, in particular means that the 
length (number of edges) of $P$ must be equal to 
$d_{G-v}(N(v))$.
In particular, as $d_{G-v}(N(v)) \leq 4$, the length of $P$ must be at most $4$. 

 If $d_{G-v}(N(v)) = 1$, then $N(v)$ is a clique.
 This implies that $v$ is a mandatory vertex with any vertex of $N(v)$ as support.

If 
$d_{G-v}(N(v)) \geq 2$, then due to the choices of $a$ and $b$, for any vertex $x \in N(v)$ we have $I(x)$ intersecting the interval $[a_r, b_l]$. 
In particular, if $d_{G-v}(N(v)) = 2$, 
then $[a_r, b_l] \subseteq I(u_1)$. Thus, $u_1$ is adjacent to all the vertices of $N(v)$,
implying $v$ is a mandatory vertex with support $u_1$. 

If $d_{G-v}(N(v)) =3$ or $4$, then we will show that $v$ is a mandatory vertex where $u_2$ is the support of $v$. Let $u_2vx$ be an induced $2$-path of $G$. Note that, 
$$[a_r, b_l] \subseteq \left( I(u_1) \cup I(u_2) \cup \cdots \cup I(u_k) \right) \cup I(x).$$
Thus, $x$ must be adjacent to, at least one of $u_1$ and $u_3$ (if it exists). Without loss of generality, suppose that $x$ is adjacent to $u_1$. Thus, the $2$-path $u_2vx$ is part of the $4$-cycle $u_2vxu_1u_2$. Therefore, we can conclude that $v$ is a 
mandatory vertex. 
\end{proof}





In order to prove the main theorem of this section, we first need to prove a crucial lemma. 

\begin{lemma}\label{left-right property P}
 There exists an interval representation of $G$ such that $\ell^*$ and $r^*$ are mandatory vertices.
\end{lemma}

\begin{proof}
 Consider any interval representation of $G$. Suppose there is no $x \in V(G)$ such that $I(x) \subseteq I(\ell^*)$, then we know that $N(\ell^*)$ is a clique. Thus, $\ell^*$ is a simplicial vertex, hence from Lemma 2.1 of~\cite{foucaud2022monitoring} and from Lemma~\ref{lem property P}, we can infer that $\ell^*$ is a mandatory vertex. Suppose on the other hand, that there are some vertices in $G$ whose corresponding interval is entirely contained in the interval of $\ell^*$. Let the set $K = \{ x \in V(G) : I(x) \subseteq I(\ell^*) \}$. As $K$ is non-empty, let $v$ be the minimal element of $K$ with respect to $<_l$. So there does not exist any vertex $t \in N(v)$, such that $\ell^*vt$ is an induced $2$-path implying $\ell^*$ is a support of $v$. Hence $v$ is a mandatory vertex. In this case, we alter the interval representation of $v$ by extending it towards the left such that $v$ becomes the minimal element of $G$ with respect to $<_l$. Notice that, this is also an interval representation of the same graph. However, in this interval representation of $G$, $v$ plays the role of $\ell^*$. 

 The proof of $r^*$ is a mandatory vertex, subject to slight modification of the interval representation of $G$, is similar. 
 \end{proof}



The main result in this section is as follows.

\begin{theorem}\label{thm interval graph}
 Let $G$ be an interval graph. Then the set of all mandatory vertices of $V(G)$ forms an MEG-set of $G$. Moreover, \textsc{MEG-set} can be solved in polynomial time on interval graphs.
\end{theorem}

\begin{proof} Let $M$ be the set of all vertices in $G$ that are \emph{mandatory vertices}. We will first prove that $M$ is an (optimal) MEG-set of $G$, and later show how to compute $M$ in polynomial time. 
That is, if $pq \in E(G)$, we have to prove that $pq$ is monitored by two vertices in $M$.
If $p, q \in M$, then the edge $pq$ is trivially monitored by the vertices $p$ and $q$. Thus, without loss of generality let us assume that $q \not\in M$. Also let us fix a particular interval representation of $G$ for which we have $l^*, r^* \in M$, for the rest of the proof. It is possible to choose and fix such an interval representation due to Lemma~\ref{left-right property P}. 

Observe that,
if $I(q) \subseteq I(u)$ for some vertex $u$, then the vertex $u \in N(q)$ is adjacent to all vertices in $N(q)$ and plays the role of a support of $q$. In particular, this would imply
that $q$ is a mandatory vertex, and thus, $q \in M$, contradicting our assumption. Therefore, there is no vertex in $G$ whose interval contains 
$I(q)$ as a sub-interval.

Let $a = p$ if $p \in M$, or $a = l^*$ if $p \not\in M$.
Let $R = \{x \in M: q <_r x\}$. Notice that, $R$ is non-empty as $r^* \in R$. 
 Moreover, let $b$ be a minimal element of $R$ with respect to $<_r$. We claim that any shortest path connecting $a$ and $b$ must have $pq$ as an edge. Let $P = a_0a_1a_2\ldots a_k$ be any shortest path connecting $a = a_0$ and $b = a_k$. To complete the proof, it is now enough to show that the edge $pq$ belongs to $P$.

Notice that $P$ must be an induced path since it is a shortest path between $a$ and $b$. 
Moreover, as 
$l_a \leq l_p \leq r_q \leq r_b$, both $I(p)$ and $I(q)$ must be contained in the interval corresponding to $P$. In particular, both $p$ and $q$ must be adjacent to some vertices of the path $P$. To be precise, $p$ (resp., $q$) must be adjacent to the vertices of $P$ having consecutive indices since $P$ is an induced path in an interval graph $G$. 

Without loss of generality assume that 
$p$ is adjacent to the vertices of the subpath 
$P_p = a_i a_{i+1} \ldots a_{i'}$ 
and $q$ is adjacent to 
$P_q = a_j a_{j+1} \ldots a_{j'}$.
Since $P_p$ and $P_q$ are shortest paths, they contain at most three vertices each (otherwise there would be a shortcut through $p$ or $q$, respectively). 
Note that the interval corresponding to 
$P_p$ (resp., $P_q$) contains $I(p)$ (resp., $I(q)$), and thus $V(P_p)$ (resp., $V(P_q)$) is adjacent to all the vertices of $N(p)$ (resp., $(N(q)$).

Next we show that the vertex $q$ belongs to the path $P$. If not, since the length of the subpath $P_q$ is $2$ or less, then $d_{G-q}(N(q)) \leq 4$, which is impossible due to Corollary~\ref{cor property P} since $q$ is not a mandatory vertex, a contradiction. 
Therefore, we can conclude that $q$ is a vertex of the path $P$.

If $p \not\in M$, then we can prove that $p$ is a vertex of $P$ in a similar way. 
On the other hand, if $p \in M$, then $a=p$, and thus, $p$ is a vertex of $P$ trivially. In either case, $pq$ is an edge of $P$. As $P$ was chosen arbitrarily, $pq$ is part of all shortest paths between $a$ and $b$, and thus it is monitored by $M$. This proves that $M$ is indeed an MEG-set of $G$.

For the moreover part, note that due to Lemma~\ref{lem property P} we know $M$ is the unique minimum MEG-set of $G$. Thus, if we can find out which vertices of $G$ belong to $M$ in polynomial time, we will be done. Observe that, using the moreover part of Corollary~\ref{cor property P}, it is possible to decide whether a given vertex of an interval graph $G$ is a mandatory vertex or not in polynomial time, by checking, for every vertex $v$ of $G$, the maximum distance within vertices of $N(v)$ in the subgraph $G-v$. 
\end{proof}

\section{Parameterized algorithms}\label{sec FPT}

We now describe two parameterized algorithms for \textsc{MEG-set}.
A parameterized problem is a decision problem with an additional integer input, denoted by $k$, known as the parameter. A parameterized problem is fixed-parameter tractable (FPT) if it can be solved in time $f(k)\cdot n^c$, where $f$ is a computable function only depending on $k$, $n$ is the size of the input, and $c$ is a constant. Further details on these notions can be found in the book~\cite{DBLP:books/sp/CyganFKLMPPS15}.

\subsection{Clique-width and diameter}

We first recall that the vertex cover number of a graph $G$ is the size of a smallest vertex cover, the treedepth is the minimal depth of a forest on the vertices of $G$ such that every pair of adjacent vertices in $G$ are descendant-ancestor, and the clique-width is the minimum number of labels needed to build $G$ with the following operations:
\begin{itemize}
 \item create a new vertex with label $i$;
 \item disjoint union of two labelled graphs;
 \item adding an edge between all vertices labelled $i$ and $j$ for $i\neq j$;
 \item relabelling all vertices with label $i$ to $j$.
\end{itemize}
All those parameters are very well-established in the study of parameterized complexity. In particular, graphs with bounded clique-width also have bounded vertex cover number and bounded tree-depth.

We next provide an algorithm using the same technique used in~\cite{KK22} for the geodetic set problem.

\begin{theorem}\label{thm:cw-diam}
 \textsc{MEG-set} is FPT when parameterized by the clique-width plus the diameter of the input graph, and thus, for parameters vertex cover number and treedepth.
\end{theorem}
\begin{proof}
 We encode the property or a set $S$ to be an MEG-set into a $MSO_1$ formula, where $MSO_1$ formulas are logical formula over vertices, edges, and set of vertices. We refer the reader to Definition 5.1.3 of~\cite{courcelle2012graph} for a more formal definition. The result will then follow from Courcelle's theorem~\cite{courcelle1990monadic}.
 \[
 \phi(S) = \forall w_1, \forall w_2, (\neg (w_1w_2\in E(G)) \lor \exists u,v,~Monitors(u, v, w_1, w_2))
 \]
 \[
 Monitors(u, v, w_1, w_2) = u\in S\land v\in S\land Visit(u, v, w_1) \land Visit(u, v, w_2)
 \]
 \begin{align*}
 Visit(u, v, w) = \bigwedge_{i\in [diam(G)-1]} & (Dist_i(u, v)\\ \land (\bigvee_{j\in [i-1]} & Dist_j(u, w) \land Dist_{i-j}(w, v)\\ &\land \nexists w' (w\neq w' \land Dist_j(u, w') \land Dist_{i-j}(w', v))))
 \end{align*}
 \begin{align*}
 Dist_i(u, w) = & \exists v_2,\dots, v_{i-1}(Path(u, v_2,\dots, v_{i-1}, w))\\ & \land \bigwedge_{j\in[i-1]}\nexists v_2,\dots, v_{j-1}(Path(u, v_2,\dots, v_{j-1}, w))
 \end{align*}
 \[
 Path(v_1, \dots, v_i) = \bigwedge_{j\in [i-1]}v_jv_{j+1}\in E(G).
 \]

The formula $Path(v_1, \dots, v_i)$ is true whenever the vertices $v_1,\ldots, v_i$ form a (not necessarily simple) path. The formula $Dist_i(u, w)$ is true whenever $u,w$ are at distance exactly $i$~\cite{KK22}. The formula $Visit(u, v, w_2)$ is true whenever $w_2$ lies on every shortest path from $u$ to $v$. $Monitors(u, v, w_1, w_2)$ is true whenever both $w_1$ and $w_2$ lie on all shortest paths from $u$ to $v$ (thus, if $w_1w_2$ is an edge, it is monitored by $u$ and $v$).
 
 The size of $\phi(S)$ is polynomial in the diameter of $G$. We prove that it is satisfied if and only if $S$ is an MEG-set of $G$, so the result follows from the extension of Courcelle's theorem to clique-width~\cite{CourcelleCW}.
 
 A set $S$ is an MEG-set if and only if every edge of $G$ is monitored by some pair of vertices of $S$. Note that $\phi(S)$ is satisfied whenever for every pair of vertices $w_1, w_2\in V(G)$, $w_1w_2$ is an edge implies that there exists $u, v$ such that $Monitors(u, v, w_1, w_2)$ is satisfied. Observe from the definition of $Monitors$ that in this case, $u, v$ are vertices of $S$, so we are left to prove that $u, v$ monitors $w_1, w_2$ if and only if $Visit(u, v, w_1)$ and $Visit(u, v, w_2)$ are satisfied. Recall that $Dist_i(u, v)$ is satisfied if and only if $u, v$ are at distance exactly $i$. Hence, $Visits(u, v, w)$ is satisfied if and only if there exists some distance $i$ where $1\leq i\leq diam(G)-1$ such that $u, v$ are at distance exactly $i$ from one another, $w$ lies on a shortest path from $u$ to $v$ and there is no $w'$ on a shortest path from $u$ to $v$ such that $d(u, w)=d(u, w')$ and $w\neq w'$. It is easy to check that the latter condition implies that every shortest path contains $w$, hence $Visit(u, v, w_1)$ and $Visit(u, v, w_2)$ are both satisfied if and only if all shortest paths contain both $w_1$ and $w_2$. Since it it a shortest path and $w_1, w_2$ are adjacent, this is equivalent to the edge $w_1w_2$ being monitored, which concludes the proof.

 Since the diameter~\cite{DBLP:journals/tcs/GimaHKKO22} and clique-width~\cite{DBLP:journals/siamcomp/CorneilR05} of a graph are both bounded above by a function of its vertex cover (and treedepth), the result for these parameters follows.
\end{proof}


\subsection{Treewidth parameterization for chordal graphs}

Next, we show that the diameter parameter from Theorem~\ref{thm:cw-diam} can be omitted, if we restrict ourselves to the weaker parameter treewidth and the class of chordal graphs. Recall that the \emph{treewidth} of a graph $G$ is the largest size of a bag (minus one) among all tree-decompositions of $G$ (see the definition below).

\begin{theorem}\label{thm:chordal}
 In chordal graphs, \textsc{MEG-set} is FPT when parameterized by the treewidth of the input graph.
\end{theorem}

We recall the (classic) definition of a \emph{nice tree decomposition} adapted to the case of chordal graphs, see~\cite{belmonte2014detecting}.

\begin{definition}
A \emph{nice tree decomposition} of a chordal graph $G$ is a rooted tree $\mathcal \mathcal{S}_{up}$ where each node $t$ is associated to a subset $X_t$ of $V(G)$ called \emph{bag}, and each internal node has one or two children, with the following properties.
\begin{enumerate}
\item The set of nodes of $\mathcal{S}_{up}$ containing a given vertex of $G$ forms a nonempty connected subtree of $\mathcal \mathcal{S}_{up}$.

\item Any two adjacent vertices of $G$ appear in a common node of $\mathcal \mathcal{S}_{up}$.

\item For each node $t$ of $\mathcal \mathcal{S}_{up}$, $G[X_t]$ is a clique.

\item Each node of $\mathcal \mathcal{S}_{up}$ belongs to one of the following types: \emph{introduce}, \emph{forget}, \emph{join} or \emph{leaf}.

\item A join node $t$ has two children $t_1$ and $t_2$ such that $X_t = X_{t_1} = X_{t_2}$.

\item An introduce node $t$ has one child $t_1$ such that $X_t \setminus \{v\} = X_{t_1}$, where $v \in X_t$.

\item A forget node $t$ has one child $t_{1}$ such that $X_t = X_{t_1} \setminus \{v\}$, where $v \in X_{t_1}$.

\item A leaf node $t$ is a leaf of $\mathcal \mathcal{S}_{up}$ with $X_t=\emptyset$.

\item The tree $\mathcal \mathcal{S}_{up}$ is rooted at a leaf node $r$ (with $X_r = \emptyset$).
\end{enumerate}
\end{definition}

Moreover, for a node $t$ of $\mathcal \mathcal{S}_{up}$, we define $G_t$ as the subgraph of $G$ induced by the vertices appearing in bags from nodes of the subtree of $\mathcal \mathcal{S}_{up}$ rooted at $t$. 
Given a chordal graph $G$ of order $n$, one can compute an optimal tree decomposition of $G$ all whose bags are cliques (i.e. where all bags are of size at most $w=tw(G)$ and with properties 1--3 above) in time $O(n)$~\cite{Blair1993}.
It can be made into a nice tree decomposition with $O(n)$ bags in time $O(wn)$~\cite{kloks1994treewidth}, and thus satisfying 1--9. 

Let $G$ be a chordal graph, with $\mathcal{T}$ a nice tree decomposition of $G$. A \emph{partial solution} for \textsc{MEG-set} on a bag $t\in \mathcal{T}$ with set of vertices $X_t$ is a set $(\mathcal{K}, \mathcal{S}_{down}, \mathcal{S}_{up}, \mathcal{C})$ where $\mathcal{K}$ is a subset of $X_t$, $\mathcal{S}_{down}$ is a subset of $\mathcal{P}(X_t)$ (where for a set $S$, $\mathcal{P}(S)$ denotes the set of all subset of $S$), $\mathcal{S}_{up}$ is a subset of $\mathcal{P}(X_t)$ and $\mathcal{C}$ is a set of pairs of an element of $X_t$ and an element of $\mathcal{P}(X_t)$. A partial solution for a node $t$ intuitively represents the behaviour of a potential MEG-set of $G$ with respect to the bag $X_t$ and the subgraph $G_t$. More precisely, $\mathcal{K}$ represents the set of vertices of the potential MEG-set in the bag $t$, $\mathcal{S}_{down}$ will be the collection of subsets of $X_t$ closest to a vertex of the partial solution in $G_t$, $\mathcal{S}_{up}$ will be the collection of subsets of $X_t$ closest to a guess of vertices of $G\setminus G_t$ completing the partial solution, and $\mathcal{C}$ is a set of constraints left by previously forgotten vertices, to be seen as the set of vertices that could become a shortcut to a shortest path needed to monitor a forgotten edge.
We call the \emph{projection} of a vertex $u$ on a set of vertices $X$ the set of vertices $\{v\in X, d(u, v) = d(u, X)\}$ where $d(u, X) = min_{v\in X}(d(u, v))$, and denote it by $proj(u, X)$.

For a node $t$ with set of vertices $X_t$, we define $opt_t(\mathcal{K}, \mathcal{S}_{down}, \mathcal{S}_{up}, \mathcal{C}) = \min(|M|)$
where $M \subseteq V(G_t)$ such that:\begin{itemize}
 \item[(P1)] $M\cap X_t = \mathcal{K}$;
 \item[(P2)] for each $S\in \mathcal{S}_{down}$, there exists $a\in M$ with $proj(a, X_t) = S$;
 \item[(P3)] for each $a\in X_t$, either $a\in \mathcal{K}$ or $\{a\}\in \mathcal{S}_{down}\cup \mathcal{S}_{up}$;
 \item[(P4)] for each edge $xy\in E(G_t)$, either $xy$ is monitored by two vertices of $M$, or $x,y\in X_t$, or there exists $S\in \mathcal{S}_{up}$ and $x'\in M$ such that for all $y'\in S$, either $\{x', y'\}$ monitors $xy$;
 \item[(P5)] for each $xy\in E(G_t)$ such that $x\in X_t$ and $y\notin X_t$, $xy$ is monitored by two vertices of $M$ or $(x, X_t\setminus (N(y)-\{x\})) \in \mathcal{C}$.
\end{itemize}

We say that a set $M$ that realizes the conditions (P1)--(P5) above \emph{realizes} the partial solution $(\mathcal{K}, \mathcal{S}_{down}, \mathcal{S}_{up}, \mathcal{C})$. If no such $M$ exists, then $opt_t(\mathcal{K}, \mathcal{S}_{down}, \mathcal{S}_{up}, \mathcal{C}) = +\infty$.

Note that $opt_t(\mathcal{K}, \mathcal{S}_{down}, \mathcal{S}_{up}, \mathcal{C})$ can indeed be equal to $+\infty$. Consider for instance a graph $G$ isomorphic to a $K_2$ with vertices $u, v$ and the nice tree decomposition $Leaf-Introduce(u)-Introduce(v)-Forget(u)-Forget(v)$. In the node $Introduce(v)$, the set of vertices of the bag is $\{u,v\}$, but any partial solution containing $\{u, v\}$ in either its set $\mathcal{S}_{down}$ or $\mathcal{S}_{up}$ will have its $opt$ equal to $+\infty$. 

\begin{lemma}\label{lemma:root}
 If $t$ is the root bag of $G$, then $opt_t(\emptyset, \emptyset, \emptyset, \emptyset)$ is the minimum size of an MEG-set of $G$. 
\end{lemma}
\begin{proof}
 By simply replacing the value $\mathcal{K}, \mathcal{S}_{down}, \mathcal{S}_{up}, \mathcal{C}$ by empty sets in the above definitions, it comes that $opt_t(\emptyset, \emptyset, \emptyset, \emptyset)$ is the minimal size of a set $M\subseteq V(G)$ such that for each $uv\in E(G_t) = G$, $uv$ is monitored by two elements of $M$, hence an MEG-set.
\end{proof}

\begin{algo}\label{algo:FPT}
 Let us consider a nice tree decomposition of $G$. Since $G$ is chordal, we can also further assume that every bag is a clique. We will produce a dynamic programming algorithm to compute every possible type of an MEG-set on the current bag, and compute the size of a smallest partial solution for every type. We treat all the bags of $G$ in time $2^{2^{O(tw)}}O(n)$.

This algorithm will compute all the types in a bottom-up fashion, where for a bag $t$, we consider any possible choice of $(\mathcal{K}, \mathcal{S}_{down}, \mathcal{S}_{up}, \mathcal{C})$ and, using the known partial solutions of child (children for a Join node), we select only the feasible solutions, as well as computing $opt_t(\mathcal{K}, \mathcal{S}_{down}, \mathcal{S}_{up}, \mathcal{C})$. 

 \begin{itemize}
 \item (Leaf) $t$ is a Leaf node. The set of partial solutions has one element, $(\emptyset, \emptyset, \emptyset, \emptyset)$ and $opt_t(\emptyset, \emptyset, \emptyset, \emptyset) = 0$.
 
 \item (Introduce) Assume $t$ is an Introduce node, adding the vertex $v$. Let us consider $\mathcal{S} = (\mathcal{K}, \mathcal{S}_{down}, \mathcal{S}_{up}, \mathcal{C})$ a partial solution of $G_t$. We denote $t'$ the child of $t$ and $X_t$ the set of vertices of $t$. There are two cases:
 \begin{itemize}
 \item If $v\in \mathcal{K}$, then $opt_t(\mathcal{K}, \mathcal{S}_{down}, \mathcal{S}_{up}, \mathcal{C}) = \min\{opt_t(\mathcal{K}', \mathcal{S}'_{down}, \mathcal{S}'_{up}, \mathcal{C}')\}+1$ where $(\mathcal{K}', \mathcal{S}'_{down}, \mathcal{S}'_{up}, \mathcal{C}')$ is any partial solution of $t'$ with:
 \begin{itemize}
 \item $\mathcal{K}' = \mathcal{K}\setminus\{v\}$;
 \item $\mathcal{S}'_{down} = \mathcal{S}_{down}$;
 \item $\mathcal{S}'_{up} = \{S\setminus \{v\}, S\in \mathcal{S}_{up}\}\cup X_t\setminus \{v\}$;
 \item $\mathcal{C}' = \mathcal{C}$.
 \end{itemize}
 
 \item else, $opt_t(\mathcal{K}, \mathcal{S}_{down}, \mathcal{S}_{up}, \mathcal{C}) = \min\{opt_t(\mathcal{K}', \mathcal{S}'_{down}, \mathcal{S}'_{up}, \mathcal{C}')\}$ where $(\mathcal{K}', \mathcal{S}'_{down}, \mathcal{S}'_{up}, \mathcal{C}')$ is any solution of $t'$ with:\begin{itemize}
 \item $\mathcal{K}' = \mathcal{K}$;
 \item $\mathcal{S}'_{down} = \{S\setminus \{v\}, S\in S\}$;
 \item $\mathcal{S}'_{up} = \{S\setminus \{v\}, S\in T\}$;
 \item $\mathcal{C}' = \mathcal{C}$.
 \end{itemize}
 In both cases, if no such $(\mathcal{K}', \mathcal{S}'_{down}, \mathcal{S}'_{up}, \mathcal{C}')$ exists, then $opt_t(\mathcal{K}, \mathcal{S}_{down}, \mathcal{S}_{up}, \mathcal{C}) = +\infty$. Otherwise, we say $(\mathcal{K}, \mathcal{S}_{down}, \mathcal{S}_{up}, \mathcal{C})$ and $(\mathcal{K}', \mathcal{S}'_{down}, \mathcal{S}'_{up}, \mathcal{C}')$ are \emph{compatible}.
 \end{itemize}
 
 \item (Forget) Assume $t$ is a Forget node of a vertex $v$, with a child $t'$ and $(\mathcal{K}, \mathcal{S}_{down}, \mathcal{S}_{up}, \mathcal{C})$ a partial solution of $t$. We denote $X_t$ the set of vertices of $t$. Then $opt_t((\mathcal{K}, \mathcal{S}_{down}, \mathcal{S}_{up}, \mathcal{C})) = \min\{opt_{t'}(\mathcal{K}', \mathcal{S}'_{down}, \mathcal{S}'_{up}, \mathcal{C}')\}$ where:\begin{itemize}
 \item $\mathcal{K}' = \mathcal{K}$ or $\mathcal{K}' = K\cup \{v\}$;
 \item If $\{v\}\in \mathcal{S}'_{down}$, then $X_t \in \mathcal{S}'_{down}$ and $\mathcal{S}'_{down} = \{S\cup \{v\}, S \in \mathcal{S}_1\}\cup \mathcal{S}_2$ for some partition of $\{S \setminus\{v\}, S \in S\setminus\{X_t\}\}$ into two disjoint sets $\mathcal{S}_1, \mathcal{S}_2$.
 Else, $\mathcal{S}'_{down} = \{S\cup \{v\}, S\in \mathcal{S}_1\}\cup \mathcal{S}_2$ for some partition of $\{S\setminus\{v\}, S\in \mathcal{S}_{down}\}$ into two disjoint sets $\mathcal{S}_1, \mathcal{S}_2$.
 \item $\mathcal{S}'_{up} = \{S\setminus \{v\}, S\in \mathcal{S}'_{up}\}$;
 \item $\mathcal{C}' = (\{(u, c\cup \{v\}), (u, c)\in \mathcal{C}_1\}\cup (\{(u, c)\in \mathcal{C}_2\})$ where $u\neq v$ for some partition of $\mathcal{C}$ into two disjoint sets $\mathcal{C}_1, \mathcal{C}_2$. 
 \end{itemize}
 
 If no such $(\mathcal{K}', \mathcal{S}'_{down}, \mathcal{S}'_{up}, \mathcal{C}')$ exists, then $opt_t(\mathcal{K}, \mathcal{S}_{down}, \mathcal{S}_{up}, \mathcal{C}) = +\infty$. Otherwise, we say $(\mathcal{K}, \mathcal{S}_{down}, \mathcal{S}_{up}, \mathcal{C})$ and $(\mathcal{K}', \mathcal{S}'_{down}, \mathcal{S}'_{up}, \mathcal{C}')$ are \emph{compatible}.
 \item (Join) $t$ is a Join node, with $t_1, t_2$ its two children. We consider $(\mathcal{K}, \mathcal{S}_{down}, \mathcal{S}_{up}, \mathcal{C})$ a partial solution of $t$. For any pair of partial solutions $\mathcal{S}_{t_1}, \mathcal{S}_{t_2}$ where $\mathcal{S}_{t_1} = (\mathcal{K}^1, \mathcal{S}^1_{down}, \mathcal{S}^1_{up}, \mathcal{C}^1)$ is a partial solution for ${t_1}$ and $\mathcal{S}_{t_2} = (\mathcal{K}^2, \mathcal{S}^2_{down}, \mathcal{S}^2_{up}, \mathcal{C}^2)$ one for ${t_2}$, we define $\mathcal{S}'$ with
 \begin{itemize}
 \item $\mathcal{K}'= \mathcal{K}^1 = \mathcal{K}^2$;
 \item $\mathcal{S}'_{down} = \mathcal{S}^1_{down} \cup \mathcal{S}^2_{down}$;
 \item $\mathcal{S}'_{up}= \mathcal{S}^1_{up}-\mathcal{S}^2_{down} = \mathcal{S}^2_{up}-\mathcal{S}^1_{down}$, where $\mathcal{S}^1_{down}\subseteq \mathcal{S}^2_{up}$ and $\mathcal{S}^2_{down}\subseteq \mathcal{S}^1_{up}$;
 \item $\mathcal{C}' = \mathcal{C}^1 \cup \mathcal{C}^2 - (\{(u, c)\in \mathcal{C}^1, \exists S \in \mathcal{S}^2_{down}, s\cap c = \emptyset \text{ and } u\in s\} \cup \{c\in \mathcal{C}^2, \exists S \in \mathcal{S}^1_{down}, s\cap c = \emptyset \text{ and } u\in s\})$. More intuitively, this is the union of the constraints that are not solved by the other child's partial solution.
 \end{itemize}
 If $(\mathcal{K}', \mathcal{S}'_{down}, \mathcal{S}'_{up}, \mathcal{C}') = (\mathcal{K}, \mathcal{S}_{down}, \mathcal{S}_{up}, \mathcal{C})$, then we say $\mathcal{S}$ is \emph{compatible} with the pair $(\mathcal{S}_{t_1}, \mathcal{S}_{t_2})$, and $opt_t(\mathcal{K}, \mathcal{S}_{down}, \mathcal{S}_{up}, \mathcal{C}) = \min\{opt_{t_1}(\mathcal{S}_{t_1}) + opt_{t_2}(\mathcal{S}_{t_2}) - |\mathcal{K}|\}$ for any pair $(\mathcal{S}_{t_1}, \mathcal{S}_{t_2})$ compatible with $\mathcal{S}$.
 If no such $(\mathcal{K}^1, \mathcal{S}^1_{down}, \mathcal{S}^1_{up}, \mathcal{C}^1)$ or $(\mathcal{K}^2, \mathcal{S}^2_{down}, \mathcal{S}^2_{up}, \mathcal{C}^2)$ exist, then $opt_t(\mathcal{K}, \mathcal{S}_{down}, \mathcal{S}_{up}, \mathcal{C}) = +\infty$.
 \end{itemize}
\end{algo}

\begin{lemma}\label{lemma:introduce}
 If $t$ is an Introduce node of a vertex $v$, with child $t'$ and $(\mathcal{K}, \mathcal{S}_{down}, \mathcal{S}_{up}, \mathcal{C})$ a partial solution of $t$, then the value computed by Algorithm \ref{algo:FPT} is correct.
\end{lemma}
\begin{proof} 
 Assume $v\in \mathcal{K}$. We prove $opt_t(\mathcal{K}, \mathcal{S}_{down}, \mathcal{S}_{up}, \mathcal{C}) = \min\{opt_{t'}(\mathcal{K}', \mathcal{S}'_{down}, \mathcal{S}'_{up}, \mathcal{C}')\}+1$ where the $min$ is taken over all compatible partial solutions $(\mathcal{K}', \mathcal{S}'_{down}, \mathcal{S}'_{up}, \mathcal{C}')$ of $t'$.
 We first note that if there exists $c\in \mathcal{C}$ with $v\in c$, or $S \in \mathcal{S}_{down}$ where $v\in S$, then no partial solution $(\mathcal{K}', \mathcal{S}'_{down}, \mathcal{S}'_{up}, \mathcal{C}')$ are compatible, and $opt_t(\mathcal{K}, \mathcal{S}_{down}, \mathcal{S}_{up}, \mathcal{C}) = +\infty$ and we are done.
 \begin{itemize}
 \item We first prove $opt_t(\mathcal{K}, \mathcal{S}_{down}, \mathcal{S}_{up}, \mathcal{C}) \leq \min\{opt_{t'}(\mathcal{K}', \mathcal{S}'_{down}, \mathcal{S}'_{up}, \mathcal{C}')\}+1$. Let us consider $M'$ a set of vertices of $V(G_{t'})$ realizing a compatible partial solution $(\mathcal{K}', \mathcal{S}'_{down}, \mathcal{S}'_{up}, \mathcal{C}')$, and such that $|M'| = \min\{opt_{t'}(\mathcal{K}', \mathcal{S}'_{down}, \mathcal{S}'_{up}, \mathcal{C}')\}$. We prove that $M = M'\cup \{v\}$ realizes $(\mathcal{K}, \mathcal{S}_{down}, \mathcal{S}_{up}, \mathcal{C})$.
 \begin{itemize}
 \item[(P1)] $M' \cap (X_t\setminus\{v\}) = \mathcal{K}'$ hence $M \cap X = \mathcal{K}$.
 \item[(P2)] Since $\mathcal{S}'_{down} = \mathcal{S}_{down}$, and $N(v)\cap V(G_t) = X_t\setminus\{v\}$, then for any $S \in S=\mathcal{S}'_{down}$, there exists $a\in M' \subset M$ such that $proj(a, X_t) = S$.
 \item[(P3)] By the same argument as before, for any vertex $a\in X_t\setminus\{v\}$, if $\{a\}\in \mathcal{S}'_{down}$ then $\{a\}\in \mathcal{S}_{down}$. Moreover, $v\in M$ thus we are left to consider the vertices $\{a\}\in X_t\setminus\{v\}$ such that $\{a\}\notin \mathcal{S}_{down}$. However, since $(\mathcal{K}, \mathcal{S}_{down}, \mathcal{S}_{up}, \mathcal{C})$ is feasible, then in fact $\{a\}\in \mathcal{S}_{up}$, and we are done.
 \item[(P4)] Note that this condition stays true for any edge not incident to $v$. Now let us consider $u'\in X_t$ such that $\{u'\}\notin \mathcal{S}'_{down}$ and $u'\notin M$, otherwise we are done. Since $(\mathcal{K}, \mathcal{S}_{down}, \mathcal{S}_{up}, \mathcal{C})$ is feasible, then either $\{u\}\in \mathcal{S}_{down}$, in which case we are done (since there exists $a\in M'$ such that $proj(a, X_t) = \{u\}$), or we know that $\{u\}\in \mathcal{S}_{up}$ holds, hence $\{u\}$ lies in $\mathcal{S}'_{up}$.
 
 \item[(P5)] Since $\mathcal{C}' = \mathcal{C}$ and all the edges of $E(G_t)$ incident to $v$ are incident to another vertex of $X_t$, this is true by $(P4)$ above.
 \end{itemize}
 \item We now prove $opt_t(\mathcal{K}, \mathcal{S}_{down}, \mathcal{S}_{up}, \mathcal{C}) \geq \min\{opt_t(\mathcal{K}', \mathcal{S}'_{down}, \mathcal{S}'_{up}, \mathcal{C}')\}+1$. Towards a contradiction, assume $opt_t(\mathcal{K}, \mathcal{S}_{down}, \mathcal{S}_{up}, \mathcal{C}) < \min\{opt_t(\mathcal{K}', \mathcal{S}'_{down}, \mathcal{S}'_{up}, \mathcal{C}')\}+1$ and consider a subset $M$ of $V(G_t)$ realizing $opt_t(\mathcal{K}, \mathcal{S}_{down}, \mathcal{S}_{up}, \mathcal{C})$. We first note that we can assume w.l.o.g. that for each $c\in \mathcal{C}$, we have $v\notin c$, because any element of $\mathcal{S}_{up}$ satisfying $c$ would also satisfy $c\setminus\{v\}$. Then, we claim that for $\mathcal{K}'' = K\setminus\{v\}$, $\mathcal{S}''_{down} = \{S \setminus \{v\}, S \in S\}$, $\mathcal{S}''_{up} = \{S\setminus \{v\}, S\in \mathcal{S}_{up}\}\cup X_t\setminus \{v\}$, and $\mathcal{C}'' = \{(c\setminus\{v\}, X), (c, X)\in C\}$, the inequality $opt_{t'}(\mathcal{K}'', \mathcal{S}''_{down}, \mathcal{S}''_{up}, \mathcal{C}'') \leq opt_t(\mathcal{K}, \mathcal{S}_{down}, \mathcal{S}_{up}, \mathcal{C})-1$ holds, which would be a contradiction since $(\mathcal{K}'', \mathcal{S}''_{down}, \mathcal{S}''_{up}, \mathcal{C}'')$ and $(\mathcal{K}, \mathcal{S}_{down}, \mathcal{S}_{up}, \mathcal{C})$ are compatible. To this end, we prove that $M\setminus\{v\}$ realizes $(\mathcal{K}'', \mathcal{S}''_{down}, \mathcal{S}''_{up}, \mathcal{C}'')$:
 \begin{itemize}
 \item[(P1)] $M\cap(X_t\setminus\{v\}) = K\setminus\{v\}$.
 \item[(P2)] For each $S \in \mathcal{S}''_{down}$, since $N(v)\cap V(G_t) = X_t\setminus\{v\}$, we have $v\notin proj(a, X_t)$ for any $a\in G_{t'}$, otherwise $opt_t(\mathcal{K}, \mathcal{S}_{down}, \mathcal{S}_{up}, \mathcal{C}) = +\infty$ and we are done. Hence, there exists $a\in M$ such that $proj(a, X_t) = S = S\setminus\{v\}$.
 \item[(P3)] For each $a\in X_t$, if there exists $S \in \mathcal{S}_{down}$ such that $S= \{a\}$, then $S \setminus\{v\} = S \in \mathcal{S}''_{down}$. If $a\in \mathcal{K}$, then in particular $\{a\}\in \mathcal{S}''_down$ and we are done as well. We now assume it is not the case, hence $\{a\}\in \mathcal{S}_{up}$. By the same argument, $\{a\}\in \mathcal{S}''_{up}$.
 \item[(P4)] Let us consider $u_1u_2\in E(G_t)$. If $u_1u_2$ is monitored by two vertices of $M\setminus\{v\}$ then we are done. Remember that if for some $S \in \mathcal{S}_{down}$, we have $v\in S$, then $opt_t(\mathcal{K}, \mathcal{S}_{down}, \mathcal{S}_{up}, \mathcal{C}) = +\infty$ and we are done. In particular, if there exists $u'\in M\setminus\{v\}$ and $S \in \mathcal{S}_{down}$ such that for all $v'\in \mathcal{S}_{down}$, $u', v'$ monitors $u_1u_2$, then this condition is still satisfied in $(\mathcal{K}'', \mathcal{S}''_{down}, \mathcal{S}''_{up}, \mathcal{C}'')$. We conclude by noticing that since $X_t$ is an induced clique and $N(v)\cap V(G_t) = X_t\setminus\{v\}$, then for any vertex $v'$ of $X_t\setminus\{v\}$, if $u', v$ monitors $u_1u_2$ then $u', v'$ also monitors $u_1u_2$ and $X_t\setminus\{v\}\in \mathcal{S}'_{down}$.
 \item[(P5)] Finally, since we can assume $\mathcal{C}'' = \mathcal{C}$ and all edges incident to $v$ are incident to another vertex of $X_t$, the last condition is fulfilled as well.
 \end{itemize}
 Since $M\setminus\{v\}$ realizes $(\mathcal{K}'', \mathcal{S}''_{down}, \mathcal{S}''_{up}, \mathcal{C}'')$, and this partial solution is compatible with $(\mathcal{K}, \mathcal{S}_{down}, \mathcal{S}_{up}, \mathcal{C})$, then $opt_{t'}(\mathcal{K}', \mathcal{S}'_{down}, \mathcal{S}'_{up}, \mathcal{C}') \leq |M|-1$ and we are done. 
 \end{itemize} 
Note that since $N(v)\cap V(G_t) = X_t\setminus\{v\}$, it cannot be that $\{v\}\in \mathcal{S}_{down}$, and the case where $v\notin \mathcal{K}$ can be solved with the same arguments as last case except using $\{v\}\in \mathcal{S}_{up}$ to monitor edges instead of $v\in \mathcal{K}$.
\end{proof}
\begin{lemma}\label{lemma:forget}
 If $t$ is a Forget node of a vertex $v$, with a child $t'$ and $(\mathcal{K}, \mathcal{S}_{down}, \mathcal{S}_{up}, \mathcal{C})$ a partial solution of $t$, then the value computed by Algorithm \ref{algo:FPT} is correct.
\end{lemma}
\begin{proof}
 Since $V(G_t) = V(G_{t'})$, we only have to prove that, if $(\mathcal{K}, \mathcal{S}_{down}, \mathcal{S}_{up}, \mathcal{C})$ and $(\mathcal{K}', \mathcal{S}'_{down}, \mathcal{S}'_{up}, \mathcal{C}')$ are compatible, then a set $M\subset V(G_t)$ realizes $(\mathcal{K}, \mathcal{S}_{down}, \mathcal{S}_{up}, \mathcal{C})$ if and only if it realizes $(\mathcal{K}', \mathcal{S}'_{down}, \mathcal{S}'_{up}, \mathcal{C}')$. In particular, the properties P1-4 are direct, and it can be verified that for any edge $wv\in E(V(G_t))$ incident to $v$, we have $X_t\setminus N(v) = \emptyset$ (since $X\cup \{v\}$ is an induced clique) and it is in $\mathcal{C}$.
 This proves that $opt_t(\mathcal{K}, \mathcal{S}_{down}, \mathcal{S}_{up}, \mathcal{C}) = \min\{opt_{t'}(\mathcal{K}', \mathcal{S}'_{down}, \mathcal{S}'_{up}, \mathcal{C}')\}$ where $(\mathcal{K}', \mathcal{S}'_{down}, \mathcal{S}'_{up}, \mathcal{C}')$ is any compatible partial solution of $t'$, which concludes the case.
\end{proof}
\begin{lemma}\label{lemma:join}
 If $t$ is a Join node of two children ${t_1}, {t_2}$, and $(\mathcal{K}, \mathcal{S}_{down}, \mathcal{S}_{up}, \mathcal{C})$ a partial solution of $t$, then the value computed by Algorithm \ref{algo:FPT} is correct.
\end{lemma}
\begin{proof}
 We prove that $opt_t(\mathcal{K}, \mathcal{S}_{down}, \mathcal{S}_{up}, \mathcal{C}) = \min\{opt_{t_1}(\mathcal{S}_{t_1}) + opt_{t_2}(\mathcal{S}_{t_2}) - |\mathcal{K}|\}$ on all pairs $(\mathcal{S}_{t_1} = (\mathcal{K}^1, \mathcal{S}^1_{down}, \mathcal{S}^1_{up}, \mathcal{C}^1), \mathcal{S}_{t_2} = (\mathcal{K}^2, \mathcal{S}^2_{down}, \mathcal{S}^2_{up}, \mathcal{C}^2))$ compatible with $\mathcal{S}$.
 
 We first show that $opt_t(\mathcal{K}, \mathcal{S}_{down}, \mathcal{S}_{up}, \mathcal{C}) \leq \min\{opt_{t_1}(\mathcal{S}_{t_1}) + opt_{t_2}(\mathcal{S}_{t_2}) - \mathcal{K}\}$. To this end, we show that if $M_1$ (resp. $M_2$) is a set of vertices realizing $\mathcal{S}_{t_1}$ (resp. $\mathcal{S}_{t_2}$) then $M_1\cup M_2$ realizes $(\mathcal{K}, \mathcal{S}_{down}, \mathcal{S}_{up}, \mathcal{C})$. We denote $X_t$ the set of vertices of $t$.

\begin{itemize}
 \item[(P1)] Since $\mathcal{K}^1 = \mathcal{K}^2 = \mathcal{K}$, we are done.
 \item[(P2)] Since $S = \mathcal{S}^1_{down}\cup \mathcal{S}^2_{down}$, we are done as well.
 \item[(P3)] Let us consider $a\in X_t$. If $\{a\}\in \mathcal{S}^1_{down}\cup \mathcal{S}^2_{down}$ or $a\in \mathcal{K}$ then we are done. We can thus assume that $a\in \mathcal{S}^1_{up}\cap \mathcal{S}^2_{up}$. Since $T = \mathcal{S}^1_{up}\cup \mathcal{S}^2_{up} - (\mathcal{S}^1_{up}\cap \mathcal{S}^2_{down}) - (\mathcal{S}^2_{up}\cap \mathcal{S}^1_{down})$ and $a\notin (\mathcal{S}^1_{down}\cup \mathcal{S}^2_{down})$, $a\in \mathcal{S}_{up}$ and then we are done.
 \item[(P4)] Since we are considering $M_1\cup M_2$, we only have to check, by symmetry, that if $uv\in E(G_{t_1})$ is an edge such that there exists $u'\in M_1$ and $S\in \mathcal{S}^1_{up}$ with, for any $v'\in S$, $\{u', v'\}$ monitors $uv$, then either:\begin{itemize}
 \item there exists $v''\in M_2$ such that $\{u', v''\}$ monitors $uv$;
 \item or there exists $S'\in \mathcal{S}_{up}$ such that, for any $v''\in S'$, $\{u', v''\}$ monitors $uv$.
 \end{itemize}
 In particular, note that the element $t$ is in $\mathcal{S}_{up}$ unless $S \in \mathcal{S}^2_{down}$, hence there exists some $b\in M_2$ such that $proj(b, X_t) = S$, in which case $\{b, u'\}$ monitors $uv$, which concludes the case.
 \item[(P5)] For an edge $uv\in E(G_t)$ where $u\in X_t$ and $v\notin X_t$, by arguments similar as in the case of $(P4)$ above, either $uv$ is monitored by a vertex of $M_1$ and all the vertices of some $S \in \mathcal{S}^2_{down}$, hence there exists $b\in M_2$ such that $t$ together with a vertex of $M_1$ monitors $uv$, or $X_t\setminus\{u\} \in \mathcal{C}$ and we are done.
\end{itemize}

 Hence, $M_1\cup M_2$ realizes $(\mathcal{K}, \mathcal{S}_{down}, \mathcal{S}_{up}, \mathcal{C})$ and $opt_t(\mathcal{K}, \mathcal{S}_{down}, \mathcal{S}_{up}, \mathcal{C}) \leq |M_1\cup M_2|= |M_1| + |M_2| - |\mathcal{K}| = \min\{opt_{t_1}(\mathcal{S}_{t_1}) + opt_{t_2}(\mathcal{S}_{t_2}) - |\mathcal{K}|\}$.

\medskip
 
 We now prove that $opt_t(\mathcal{K}, \mathcal{S}_{down}, \mathcal{S}_{up}, \mathcal{C}) \geq \min\{opt_{t_1}(\mathcal{S}_{t_1}) + opt_{t_2}(\mathcal{S}_{t_2}) - |\mathcal{K}|\}$.

 We consider a set $M$ of vertices of $V(G_{t})$ realizing $opt_t(\mathcal{K}, \mathcal{S}_{down}, \mathcal{S}_{up}, \mathcal{C})$, and show that there exists some partial solutions$(\mathcal{K}^1, \mathcal{S}^1_{down}, \mathcal{S}^1_{up}, \mathcal{C}^1)$ of $t_1$, and $(\mathcal{K}^2, \mathcal{S}^2_{down}, \mathcal{S}^2_{up}, \mathcal{C}^2)$ of $t_2$ such that $M_1 = M\cap V(G_{t_1})$ (resp. $M_2 = M\cap V(G_{t_2}$) realizes $(\mathcal{K}^1, \mathcal{S}^1_{down}, \mathcal{S}^1_{up}, \mathcal{C}^1)$ (resp. $(\mathcal{K}^2, \mathcal{S}^2_{down}, \mathcal{S}^2_{up}, \mathcal{C}^2))$). We will show it only for $t_1$, as the case is completely symmetrical. We define $\mathcal{S}_{t_1}$ as follows:
 \begin{itemize}
 \item $\mathcal{K}^1 = \mathcal{K}$;
 \item $\mathcal{S}^1_{down} = \{proj(a, X_t), a\in M_1$\};
 \item $\mathcal{S}^1_{up} = \mathcal{S}^2_{down}\cup \mathcal{S}_{up}$;
 \item $\mathcal{C}^1 = C \cup \{(v, X_t\setminus\{v\}, \{v\}\in \mathcal{S}^2_{down}\}$.
 \end{itemize}
 We now prove that $M_1$ realizes $(\mathcal{K}^1, \mathcal{S}^1_{down}, \mathcal{S}^1_{up}, \mathcal{C}^1)$.
 \begin{itemize}
 \item[(P1)] It is direct to check that $M\cap V(G_{t_1})\cap X_t = \mathcal{K}^1$.
 \item[(P2)] We consider $S \in \mathcal{S}^1_{down}$. By choice of $\mathcal{S}^1_{down}$, there exists $a\in M_1$ such that $proj(a, X_t) = S$.
 \item[(P3)] Let us consider $a\in X_t$. If $a\in \mathcal{K}$, then we are done. If $a\in \mathcal{S}_{up}$, then $a\in \mathcal{S}^1_{up}\setminus \mathcal{S}^2_{down}$ which is a subset of $\mathcal{S}^1_{up}$. Finally, if $a\in \mathcal{S}_{down}$, and since $S=\mathcal{S}^1_{down}\cup \mathcal{S}^2_{down}$ where $\mathcal{S}^2_{down}\subseteq \mathcal{S}^1_{up}$, then $a\in \mathcal{S}^1_{down}\cup \mathcal{S}^1_{up}$.
 \item[(P4)] We consider $uv\in E(G_{t_1})$. Since $T\subseteq \mathcal{S}^1_{up}$, if $uv$ is monitored by a vertex of $M$ and all the vertices of some $S \in \mathcal{S}_{up}$, or by two vertices of $M$, we can assume $uv$ is monitored by a vertex of $M\cap V(G_{t_2})$ and either another vertex of $M\cap V(G_{t_2})$, all the elements of some $S \in \mathcal{S}_{up}$ or a vertex of $M\cap V(G_{t_1})$ (otherwise we are done). Assume $uv$ is monitored by two vertices of $M\cap V(G_{t_2})$, denoted $a', b'$. In particular, $proj(a', X_t)\cap proj(b', X_t)=\emptyset$ --- otherwise any shared vertex of $X_t$ would create a shortcut. Since $G$ is chordal, it must be that both $u, v\in X_t$ and we are done. If $uv$ is monitored by a vertex $a$ of $M\cap V(G_{t_1})$ and a vertex $b$ of $M\cap V(G_{t_2})$, then by a similar argument, there exists $S \in \mathcal{S}^1_{up}$ such that for all $b'\in S$, $a, b'$ monitor $uv$. Finally, if $uv$ is monitored by some $b$ of $M\cap V(G_{t_2})$ and any of the elements of some $S \in \mathcal{S}_{up}$, by similar arguments, $u,v\in X_t$. 
 \item[(P5)] Let us consider two adjacent vertices $u\in X_t$ and $v\in N(v)\cap V(G_{t_1})$. If $uv$ is monitored by two vertices of $M_1$, then we are done. By argument similar as before, $uv$ cannot be monitored by two vertices of $M_2$ (otherwise $v\in X_t$). Hence, $uv$ is monitored by some $a\in M_1$ and some $b\in M_2$. Since $u\in X_t$, then $proj(b, X_t) = \{u\}$ and $\{u\}\in \mathcal{S}^2_{down}$, hence $(u, X_t\setminus\{u\})\in \mathcal{C}^1$, which concludes the proof.
 \end{itemize}
 Since there exist some $\mathcal{S}_{t_1}, \mathcal{S}_{t_2}$ compatible with $(\mathcal{K}, \mathcal{S}_{down}, \mathcal{S}_{up}, \mathcal{C})$, where $opt_t(\mathcal{K}, \mathcal{S}_{down}, \mathcal{S}_{up}, \mathcal{C}) \geq opt_{t_1}(\mathcal{S}_{t_1}) + opt_{t_2}(\mathcal{S}_{t_2}) - |\mathcal{K}|$, then in particular $opt_t(\mathcal{K}, \mathcal{S}_{down}, \mathcal{S}_{up}, \mathcal{C})$ is bigger than the minimum value over all compatible pairs.
\end{proof}
One can easily check that the function $opt_t$ can be computed in time $O(f(tw))$, since each type of node requires to compute all the possible partial solutions of the bag at hand and its child (children for Join nodes). Dealing with all sets of elements of a bag can be done in a time $2^{2^{O(tw)}}$, which is the order of the number of partial solutions of a bag and its child (squared for the children of a Join node). Since there are $O(n)$ bags, this algorithm runs in time $2^{2^{O(tw)}}n$.

\begin{proof}[Proof of Theorem~\ref{thm:chordal}]
 We provide an algorithm from Algorithm \ref{algo:FPT}, and the correctness is ensured by Lemmas \ref{lemma:introduce}, \ref{lemma:forget} and \ref{lemma:join} to prove this result.
\end{proof}

\section{Approximation algorithm}\label{sec:approx}

We now provide our approximation algorithm for \textsc{MEG-set}. 
An $\alpha$-approximation algorithm for a minimization problem is an algorithm which returns a solution with cost at most $\alpha$ times the cost of a minimum solution, where $\alpha$ is known as the approximation factor of the algorithm. Further details on these notions can be found in the book~\cite{DBLP:books/daglib/0004338}.
Our algorithm has an approximation factor that is at most $(\ln m - \ln\ln m + 0.78)\cdot (OPT-1)$ and $\sqrt{n\ln m }$, where $n$ is the number of vertices, $m$ is the number of edges, and $OPT$ is the size of a minimum MEG-set of the input graph. Note that since $OPT$ is in the approximation factor and we always have $OPT\leq n$, the approximation factor $(\ln m - \ln\ln m + 0.78)\cdot (OPT-1)$ is nontrivial only when $\ln m\cdot OPT<n$. 

The algorithm composes a reduction to \textsc{Set Cover} and an approximation algorithm for \textsc{Set Cover}.

\begin{mdframed}
\textsc{Set Cover} \\
\textbf{Instance:} A set $U$ of $m$ elements and a set $\mathcal{S}$ of $n$ subsets of $U$ such that every element of $U$ is in some set in $\mathcal{S}$. \\
\textbf{Question:} Find a minimum cardinality subset $\mathcal{S}^*$ of $\mathcal{S}$ such that every element in $U$ is in some set in $\mathcal{S}^*$. 
\end{mdframed}

There is a simple greedy algorithm for \textsc{Set Cover} - keep choosing a set in $\mathcal{S}$ which contains a maximum number of uncovered elements, and repeat until all elements are covered. This algorithm achieves an approximation ratio of $\ln m - \ln\ln m + 0.78$~\cite{DBLP:journals/jal/Slavik97}. Dinur and Steurer~\cite{DBLP:conf/stoc/DinurS14} have proved that this is almost tight. They proved that 
\textsc{Set Cover} does not admit a $(1-o(1))\ln m$-approximation algorithm, assuming P $\neq$ NP. Our algorithm for \textsc{MEG-set} is given below.

\begin{mdframed}
 \textbf{Algorithm 1}\\
 \textbf{Input:} A graph $G$.\\
 \textbf{Output:} An MEG-set for $G$.
 \begin{description}
 \item[Step 1:] For every unordered pair of vertices $u,v$ of $G$, compute $S_{\{u,v\}}$, the set of edges monitored by $\{u,v\}$. Let $\mathcal{S}$ denote the union of all these sets.
 \item[Step 2:] Let $U=E(G)$. Apply the greedy algorithm for \textsc{Set Cover} on $(U, \mathcal{S})$ to obtain a solution $\mathcal{\hat{S}}$.
 \item[Step 3:] Return $M = \bigcup_{S_{\{u,v\}}\in \mathcal{\hat{S}}}\{u,v\}$.
 \end{description}
\end{mdframed}

\begin{theorem}
 \label{thm:approx}
 Algorithm 1 is an approximation algorithm for \textsc{MEG-set} with factor $(\ln m - \ln\ln m + 0.78)\cdot (OPT-1)$ and $\sqrt{n\ln m }$, where $n$ is the number of vertices and $m$ is the number of edges, and $OPT$ is the size of a minimum MEG-set of the input graph.
\end{theorem}

\begin{proof}
 As it is easy to compute the set of edges monitored
 by a pair of vertices (by running a breadth-first search at one of the vertices, one can compute the set $S$ of edges that are in a shortest path between these vertices; the monitored edges are those that form a cut-edge in the subgraph induced by $S$), the algorithm runs in polynomial-time. Further, it is straight-forward to verify that $M$ is an MEG-set of $G$.
 Let $\alpha = \ln m - \ln\ln m + 0.78$.
 By $\mathcal{S}^*$ we denote a minimum set cover
 for $(U, \mathcal{S})$, and by $M^*$ we denote a minimum MEG-set for $G$. It is clear from Step 3 of the algorithm that $|M|\leq 2|\mathcal{\hat{S}}|$. By the approximation ratio achieved by the greedy algorithm for \textsc{Set Cover}, we obtain that $|\mathcal{\hat{S}}|\leq \alpha\cdot |\mathcal{S}^*|$. Since ${|M^*| \choose 2}$ pairs of vertices cover all edges of $G$, we obtain that $|\mathcal{S}^*|\leq |M^*|(|M^*|-1)/2 = OPT(OPT-1)/2$. It is sufficient to prove that
 $|M|/{OPT}\leq \alpha (OPT-1)$.
 \begin{equation*} \label{eq2}
 \begin{split}
 \frac{|M|}{OPT} & \leq \frac{2|\mathcal{\hat{S}}|}{OPT} \\
 & \leq \frac{2\alpha|\mathcal{S}^*|}{OPT}\\
 & \leq \frac{2\alpha\cdot OPT(OPT-1)}{2\cdot OPT}\\
 & = \alpha (OPT-1)\ .
 \end{split}
 \end{equation*}

For the second approximation ratio, if $OPT<\sqrt{\frac{n}{\ln m}}$, the previous equation implies that the above algorithm yields an approximation factor of at most:

 \begin{equation*} \label{eq4}
 \begin{split}
 \frac{|M|}{OPT} & \leq \alpha (OPT-1) \\
 & < \ln m \sqrt{\frac{n}{\ln m}}\\
 & = \sqrt{n\ln m}\ .
 \end{split}
 \end{equation*}

On the other hand, we also have $|M|\leq n$. Thus, if $OPT\geq\sqrt{\frac{n}{\ln m}}$, we have an approximation ratio of:
 \begin{equation*} \label{eq4}
 \begin{split}
 \frac{|M|}{OPT} & \leq \frac{n}{\sqrt{\frac{n}{\ln m}}} \\
 & \leq \sqrt{n\ln m}\ .
 \end{split}
 \end{equation*}
 
This completes the proof. 
\end{proof}

\section{Hardness results}\label{sec complexity}

In this section, with a simple reduction from \textsc{Vertex Cover}, we prove that there exists a constant $\alpha$ such that there is no $\alpha$-approximation algorithm for the computational problem \textsc{MEG-set}, even for 4-degenerate graphs, assuming P $\neq$ NP. 
In addition to this, we prove that the problem cannot be solved in subexponential-time, assuming the Exponential Time Hypothesis. Bil\`{o} et al.~\cite{bilo2024inapproximability} recently proved that the problem does not admit any $(c\log n)$-approximation algorithm (for any constant $c < 1/2$), assuming P $\neq$ NP. We note that the graphs they obtain in their reduction can be quite dense. Our inapproximability result, though weaker, applies to sparse graphs.

We complement these results by obtaining a $(\ln m \cdot OPT)$-approximation algorithm for \textsc{MEG-set}, where $m$ is the number of edges and $OPT$ is the size of a minimum MEG-set of the input graph. 
By $vc(G)$, we denote the size of a minimum vertex cover of $G$. We recall the computational decision problem \textsc{Vertex Cover} in the following.

\begin{mdframed}
\textsc{Vertex Cover} \\
\textbf{Instance:} A graph $G$, an integer $k$. \\
\textbf{Question:} Is it true that $vc(G)\leq k$?
\end{mdframed}

We give a reduction from \textsc{Vertex Cover} to \textsc{MEG-set}. Let $(G,k)$ be an instance of \textsc{Vertex Cover}. Without loss of generality, we assume that $G$ has at least two vertices. We apply the following reduction to get an instance $(\hat{G}, k)$ of \textsc{MEG-set}.

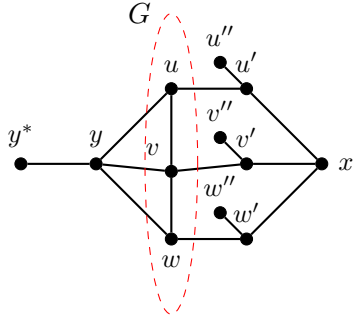
\begin{figure}
	\centering
	\begin{tikzpicture}[myv/.style={circle, fill=black, draw, inner sep=1.6pt}]
		\node[label={180:$\huge{G}$}] (gl) at (0,2) {};
		\node[myv,label={105:\small $v$}] (v) at (0,-0.1) {};
		\node[myv,label={90:\small $u$}] (u) at (0,1) {};
		\node[myv,label={270:\small $w$}] (w) at (0,-1) {};
		\node[myv,label={90:\small $v'$}] (vd) at (1,0) {};
		\node[myv,label={90:\small $u'$}] (ud) at (1,1) {};
		\node[myv,label={90:\small $w'$}] (wd) at (1,-1) {};
		\node[myv,label={0:\small $x$}] (x) at (2,0) {};
		\node[myv,label={90:\small $y$}] (y) at (-1,0) {};
		\node[myv,label={90:\small $y^*$}] (ys) at (-2,0) {};
		\node[myv,label={90:\small $v''$}] (vdd) at (0.65,0.35) {};
		\node[myv,label={90:\small $u''$}] (udd) at (0.65,1.35) {};
		\node[myv,label={90:\small $w''$}] (wdd) at (0.65,-0.65) {};
		\draw[thick] (ys) -- (y);    
		\draw [thick] (y) -- (u);    
		\draw [thick] (y) -- (v);    
		\draw [thick](y) -- (w);    
		\draw [thick](u) -- (v);    
		\draw [thick](v) -- (w);    
		\draw [thick](u) -- (ud);    
		\draw [thick](ud) -- (udd);    
		\draw [thick](v) -- (vd);    
		\draw [thick](vd) -- (vdd);    
		\draw [thick](w) -- (wd);    
		\draw [thick](wd) -- (wdd);    
		\draw[thick] (ud) -- (x);    
		\draw[thick] (vd) -- (x);    
		\draw [thick](wd) -- (x);
		\draw [dashed, red] (0,0) ellipse (0.35 and 2);
	\end{tikzpicture}
	\caption{Construction of $\hat{G}$ from $G = P_3$ as explained in the reduction.}
	\label{fig:red}
\end{figure}

\vspace{5pt}
\noindent\textbf{Reduction:}
Given a graph $G$ with at least two vertices, we build a graph $\hat{G}$ as follows. Take a copy of $G$. Let $U$ denote the set of vertices of the copy of $G$. For every vertex $u\in U$, introduce two new vertices $u'$ and $u''$ and add the edges $uu'$ and $u'u''$. Let $U'$ and $U''$ denote the set of all $u'$s and $u''$s respectively, i.e., $U'= \{u'~|~ u\in U\}$, and $U''= \{u''~|~ u\in U\}$. Further, introduce three special vertices
$x, y, y^*$, and make $x$ adjacent to all vertices in $U'$, make $y$ adjacent to all vertices in $U$, and add an edge $yy^*$. This completes the reduction. See Figure~\ref{fig:red} for an illustration.

\begin{lemma}
 \label{lem:red:forward}
 Let $C$ be a vertex cover of $G$. Then $M = C\cup U'' \cup \{y^*\}$ is an MEG-set for $\hat{G}$.
\end{lemma}
\begin{proof}
Let $u,v\in U$ where $u$ and $v$ are adjacent. Without loss of generality, assume that $u\in C$. Then $uv$ is monitored by the pair $u, v''\in M$.
For any vertex $u\in U$, the edges $yu, uu',$ and $u'u''$ are monitored by $y^*, u''\in M$.
The edge $u'x$ is monitored by $u'', v''\in M$. The edge $yy^*$ is monitored by the pair $y^*, u''$. Therefore, for every edge $e$ of $\hat{G}$, there is a pair of vertices in $M$ monitoring $e$. 
\end{proof}

Before inspecting the backward direction of the reduction, we need the following lemma.

\begin{lemma}
 \label{lem:red:backward-obs}
 Let $a,b$ be any two vertices in $\hat{G}$.
 Let $u,v\in U$ be such that $u$ is adjacent to $v$ and $\{a,b\}\cap \{u,v\}=\emptyset$. 
 Then there is at least one shortest path between $a$ and $b$ which does not contain $uv$.
\end{lemma}
\begin{proof}
 The observation is trivial if $a$ and $b$ are adjacent. Therefore, assume that $a$ and $b$ are not adjacent.
 
 Let $a,b\in U'\cup U''\cup \{x\}$. We claim that there is a 
 unique shortest path between them and the shortest path contains only vertices from $U'\cup U''\cup \{x\}$. If $a=x$ and $b=u''\in U''$, then the shortest path between them is $xu'u''$. 
 If $a=u'\in U'$ and $b=v'\in U'$, then the shortest path between them is $u'xv'$. If $a=u'\in U'$ and $b=v''\in U''$, then the shortest path between them is $u'xv'v''$. 

 It is straight-forward to verify that if $a\in \{y,y^*\}$ and $b\in U'\cup U''$, 
 then there is a unique shortest path between them (through the vertex in $U$ corresponding to $b$) and it does not contain the edge $uv$.
 Similarly, if $a\in \{y,y^*\}$ and $b=x$,
 then there are exactly $|U|$ shortest paths between them (through each vertex in $U$) and none of them contains the edge $uv$.
 It can be also seen that if $a\in \{y,y^*\}$
 and $b\in U$, then there is a unique shortest path between them and it does not contain $uv$. 

 Now, assume that $a\in U$ and $b\in U'\cup U''\cup \{x\}$. Let $b=w'\in U'$. If $w$ is not adjacent 
 to $a$, then there is a shortest path $aa'xw'$ not
 containing the edge $uv$. If $w$ is adjacent to $a$, then the unique shortest path between them, $aww'$, does not contain $uv$. The case when $b\in U''$ can be proved in a similar way. If $b=x$,
 then the unique shortest path is $aa'x$.

 The only case remaining is when $a,b\in U$.
 Since $\{a,b\}$ is disjoint with $\{u,v\}$,
 if there is a two-path between $a$ and $b$,
 through another vertex in $U$, then the path 
 does not contain $uv$. Otherwise, the unique 
 shortest path between them is $ayb$.
\end{proof}

\begin{lemma}
 \label{lem:red:backward}
 Let $M$ be an MEG-set of $\hat{G}$. 
 Then $C = M\cap U$ is a vertex cover of $G$.
\end{lemma}
\begin{proof}
 Let $uv$ be an edge in $G$. By Lemma~\ref{lem:red:backward-obs}, for every
 pair of vertices $a,b$ in $\hat{G}$
 such that $\{a,b\}\cap \{u,v\}=\emptyset$,
 there is a 
 shortest path between $a$ and $b$ not containing $uv$. Therefore, to monitor $uv$ either $u$ or 
 $v$ must be in $M$. This implies that $C$ is a 
 vertex cover of $G$.
\end{proof}

\begin{lemma}
 \label{lem:red}
 $G$ has a vertex cover of size at most $k$ if and only if $\hat{G}$ has an MEG-set of size at most 
 $k+n+1$.
\end{lemma}
\begin{proof}
 The forward direction is implied by Lemma~\ref{lem:red:forward}. For the backward
 direction, assume that $\hat{G}$ has an MEG-set
 $M$ of size at most $k+n+1$. Since every pendant vertex must be in $M$, we obtain that $U''\cup \{y^*\}\subseteq M$. 
 Then, $M\cap U$, which is of size at most $k$, is a vertex cover of $G$ by Lemma~\ref{lem:red:backward}.
\end{proof}

It is known that \textsc{Vertex Cover} is NP-complete
even for 2-degenerate planar graphs~\cite{DBLP:journals/jct/Mohar01,yannakakis1981edge}. Recall that, a graph is $k$-degenerate if every subgraph of it has a vertex of degree at most $k$ and a graph is $k$-apex if it contains a set of at most $k$ vertices whose removal yields a planar graph. Observe that, if $G$ is a 2-degenerate planar graph, then $\hat{G}$ is a 3-degenerate 2-apex graph.

\begin{corollary}
 \label{cor:npc}
 \textsc{MEG-set} is NP-complete even for 3-degenerate 2-apex graphs.
\end{corollary}

The Exponential Time Hypothesis (ETH) essentially says that \textsc{3-SAT} cannot be solved in time $2^{o(n)}$-time, where $n$ is the number of variables in the \textsc{3-SAT} instance. The sparsification lemma by Impagliazzo, Paturi, and Zane~\cite{DBLP:journals/jcss/ImpagliazzoPZ01} implies that, assuming ETH, \textsc{3-SAT} cannot be solved in time $2^{o(n+m)}$-time, where $m$ is the number of clauses in the \textsc{3-SAT} instance. In order to transfer this complexity lower bound to other problems, it is sufficient to design a polynomial-time reduction in which the size of the resultant instance is linear in the size of the input instance. We refer to Chapter 14 of \cite{DBLP:books/sp/CyganFKLMPPS15} for a detailed exposition of these concepts.

The standard reductions from \textsc{3-SAT} to \textsc{Vertex Cover}~\cite{DBLP:books/fm/GareyJ79,yannakakis1981edge} imply that, assuming ETH, \textsc{Vertex Cover} cannot be solved in time $2^{o(n+m)}$-time even for 2-degenerate graphs, where $n$ and $m$ are the number of vertices and the number of edges of the input graph.
We observe that, given a graph $G$ with $n$ vertices and $m$ edges, our reduction gives a graph $\hat{G}$
with $3n+3$ vertices and $m+4n+1$ edges.
This gives our next corollary.

\begin{corollary}
 \label{cor:eth}
 Assuming the ETH, \textsc{MEG-set} cannot be solved in time $2^{o(n+m)}$-time, even for 3-degenerate graphs.
\end{corollary}

Now, let us consider the optimization version of \textsc{MEG-set}: given a graph $G$, find the minimum-sized MEG-set of $G$. For convenience, we use the same name for both decision and optimization versions of problems - the meaning will be clear from the context.
An $\alpha$-approximation algorithm, for a parameter $\alpha > 1$, for a minimization problem $Q$ is an algorithm running in polynomial-time which gives a solution of cost at most $\alpha$-times the cost of an optimum solution for $Q$. 
A polynomial-time approximation algorithm (PTAS) for $Q$ is a polynomial-time algorithm which takes as input an instance of $Q$ and a parameter $\epsilon > 0$, and produces a solution of cost at most $(1+\epsilon)$-times the cost of an optimum solution for the instance. 

Alimonti and Kann~\cite{DBLP:journals/tcs/AlimontiK00} have proved that \textsc{Vertex Cover} does not admit a PTAS, assuming P $\neq$ NP, even for cubic graphs. We apply the reduction on a cubic graph $G$ and obtain $\hat{G}$. Without loss of generality, we can assume that $G$ is non-bipartite as \textsc{Vertex Cover} is solvable in polynomial-time for bipartite graphs. The following observation comes handy in our inapproximability results. 

\begin{observation}
 \label{obs:noptasobs}
 Let $G$ be a non-bipartite cubic graph. 
 Then $vc(G)\geq n/2+1$, where $n$ is the number of vertices in $G$.
\end{observation}
\begin{proof}
 Since $G$ is cubic, $G$ has $3n/2$ edges.
 Since a vertex can cover only 3 edges, we obtain 
 that $vc(G)\geq n/2$. This becomes an equality when
 each edge is covered uniquely by a vertex in a minimum vertex cover. Then, the vertex cover is an independent set, and hence $G$ is a bipartite graph, which is a contradiction. Therefore, $vc(G)\geq n/2+1$.
\end{proof}

\begin{theorem}
 \label{thm:noptas}
 Assuming P $\neq$ NP, \textsc{MEG-set} does not admit a polynomial-time approximation scheme, even for 4-degenerate graphs.
\end{theorem}
\begin{proof}
 Since $G$ is cubic, we obtain that $\hat{G}$ is 4-degenerate.
 
 Let $\epsilon > 0$ be any constant. Let $\alpha = 1+\epsilon/3$. Assume that there is an $\alpha$-approximation algorithm $A$ for \textsc{MEG-set}. 
 We will obtain a $(1+\epsilon)$-approximation 
 algorithm for \textsc{Vertex Cover}, which is a contradiction.

 Let $M$ be the solution returned by $A$ on $\hat{G}$. Let $\varphi(\hat{G})$ denote the size of a minimum MEG-set of $\hat{G}$. Let $C$ be $M\cap U$.

 \begin{equation*} \label{eq1}
 \begin{split}
 \frac{|C|}{vc(G)} & = \frac{|M|- (n+1)}{vc(G)} \\
 & \leq \frac{\alpha \varphi(\hat{G}) - (n+1)}{vc(G)}\ \text{(as $A$ is an $\alpha$-approximation algorithm)} \\
 & = \frac{\alpha (vc(G)+(n+1)) - (n+1)}{vc(G)}\ \text{(by Lemma~\ref{lem:red})}\\
 & = \alpha + \frac{(\alpha-1)(n+1)}{vc(G)}\\
 & \leq \alpha + \frac{(\alpha-1)(n+1)}{n/2+1}\ \text{(by Observation~\ref{obs:noptasobs})}\\
 & \leq \alpha + 2(\alpha-1)\\
 & = 1 + \frac{\epsilon}{3} + 2\frac{\epsilon}{3}\\
 & = 1+\epsilon
 \end{split}
 \end{equation*}
 This completes the proof. 
\end{proof}

A problem $Q$
is in APX if there is a constant $\alpha$ such that
there is an $\alpha$-approximation algorithm for $Q$. The problem $Q$ is APX-hard, 
if there is a PTAS-reduction from every problem in APX to $Q$. Similar to proving NP-hardness, to show that $Q$ is APX-hard, it is sufficient to show a PTAS-reduction from an APX-hard problem to $Q$. The essential requirement of a PTAS-reduction is the existence of a computable function $f$ such that 
$|M|/\varphi(\hat{G}) \leq f(1+\epsilon)\Rightarrow |C|/vc(G)\leq 1+\epsilon$. This is what we proved in Theorem~\ref{thm:noptas}.
We refer to \cite{DBLP:conf/coco/Crescenzi97} for a short guide to these topics.
Since our reduction qualifies as a PTAS-reduction and \textsc{Vertex Cover} is APX-hard even for cubic non-bipartite graphs~\cite{DBLP:journals/tcs/AlimontiK00}, we obtain the following stronger statement.

\begin{corollary}
 \label{cor:apx-hard}
 \textsc{MEG-set} is APX-hard, even for 4-degenerate graphs.
\end{corollary}

\section{Concluding remarks}\label{sec conclusions}

In this paper, we focus on the algorithmic results for the parameter, monitoring edge-geodetic number. We end the paper with a brief summary and future direction of research related to each section. 

 We showed in Section~\ref{sec:interval}, that {\sc MEG-set} is polynomial-time solvable on interval graphs, and in Section~\ref{sec FPT}, that it is FPT for chordal graphs when parameterized by the treewidth. Thus, it will be interesting to study if \textsc{MEG-set} is polynomial-time solvable on chordal graphs. In Section~\ref{sec:approx} we have given an algorithm with factor $\sqrt{n\ln m}$. It is recently shown in~\cite{bilo2024inapproximability} that \textsc{MEG-set} is not approximable in polynomial time within a factor of $0.5\ln n$. A question in this direction is to find the optimal approximation complexity of {\sc MEG-set}. Finding parameterized complexity of {\sc MEG-set} for standard parameters like the solution size, the treewidth, or the feedback edge set number can be considered subsequently. In Section~\ref{sec complexity}, we have proved that {\sc MEG-set} is NP-complete even for $2$-apex graphs. Hence, a natural question is to find the computational complexity of {\sc MEG-set} for planar graphs.

\medskip

\noindent \textbf{Acknowledgements:} This work is partially supported 
by the following projects: IFCAM (MA/IFCAM/18/39), SERB-MATRICS (MTR/2021/000858 and MTR/2022/000692), French government IDEX-ISITE initiative 16-IDEX-0001 (CAP 20-25), International Research Center "Innovation Transportation and Production Systems" of the I-SITE CAP 20-25, ANR project GRALMECO (ANR-21-CE48-0004), and Centro de Modelamiento Matemático (CMM) BASAL fund FB210005 for center of excellence from ANID-Chile.

\bibliographystyle{abbrv}
\bibliography{reference_simple_}

\end{document}